\documentclass[twocolumn,twoside]{article}
\usepackage{graphics}
\newcommand{\kms}{$\mathrm{\,km\,s}^{-1}$ }

\newcommand{\hb}{\\ \hspace*{2ex}}

\begin{document}
\title{H$_{\alpha}$ LINE AS AN INDICATOR OF ENVELOPE PRESENCE AROUND THE
CEPHEID POLARIS Aa ($\alpha~ UMi$).}
\author{I.A.\,Usenko$^1$, A.S.\,Miroshnichenko$^2$,
 V.G.\,Klochkova$^3$, N.S.\,Tavolzhanskaya$^3$\\[2mm]
\begin{tabular}{l}
 $^1$ Astronomical Observatory, Odessa National University\hb
 T.G. Shevchenko Park, Odessa 65014, Ukraine {\em igus@deneb1.odessa.ua}\\
$^2$ Department of Physics and Astronomy, University of North Carolina at Greensboro,\hb
Greensboro, NC 27402-6170, USA {\em a$\_$mirosh@uncg.edu}\\ $^3$
Special Astrophysical Observatory, Russian Academy of Sciences\hb
 Nizhnij Arkhyz, Karachaevo-Cherkessia, 369167 Russia {\em valenta@sao.ru;
 panch@sao.ru}\\[2mm]
\end{tabular}
}
\date{}
\maketitle
\indent
ABSTRACT.

We present the results of the radial velocity ($RV$) measurements of metallic
lines as well as H$_{\alpha}$ (H$_{\beta}$) obtained in 55
high-resolution spectra of the Cepheid $\alpha$ UMi (Polaris Aa) in
1994--2010. While the $RV$ amplitudes of these lines are roughly equal, their mean $RV$
begin to differ essentially with growth of the Polaris Aa pulsational activity. This
difference is accompanied by the H$_{\alpha}$ line core asymmetries on the red side mainly
(so-called knife-like profiles) and reaches 8--12 \kms
in 2003 with a subsequent decrease to 1.5--2 \kms. We interpret a so
unusual behaviour of the H$_{\alpha}$ line core as
dynamical changes in the envelope around Polaris Aa. \\[1mm]


{\bf 1. Introduction}\\[1mm]

Detection of an extended envelope around the Cepheid Polaris (hereafter
Polaris Aa) using a near-infrared interferometer (M\'{e}rand et al. 2006) suggested
an idea to check its presence spectroscopically. Usenko et al. (2013,
2014ab), Usenko and Klochkova (2015) revealed that the H$_{\alpha}$ absorption line
could be used as an indicator of the envelope presence not only in long-period
Cepheids but also in short-period ones. As a rule, Cepheids with pulsational
periods longer than 7--10$^d$ demonstrate a pronounced appearance of
a secondary variable absorption in the H$_{\alpha}$ line cores, while short-period ones
exhibit a smoother, so called knife-like shape. Besides that, a slight change
in the $RV$ of the H$_{\alpha}$ line core with pulsational phase compared
to that determined from the metallic lines is another indicator of the envelope
presence in Cepheids.

Hence the main goal of this work is to measure the $RV$s of Polaris Aa
in different pulsational phases using the metallic lines and H$_{\alpha}$
(in some cases H$_{\beta}$) line cores and to estimate visually the
shape of the latter ones.\\

{\bf 2. Observations}\\[1mm]

Observations of Polaris Aa have been obtained using the following facilities:

\begin{enumerate}
\item 1\,m telescope of the Ritter Observatory, University of Toledo (Ohio,
USA) - fiberfed echelle spectrograph 1150$\times$1150 pixel CCD
($\lambda \lambda$ 5800--6800 \AA).
\item 2.1\,m Otto Struve telescope of the McDonald Observatory (Texas, USA) - SANDIFORD
spectrograph (McCarthy et al. 1993) 1200$\times$400 pixel CCD ($\lambda \lambda$
5500--7000 \AA).
\item 6\,m telescope BTA - SAO RAS (Russia) - LYNX (Panchuk et al. 1993), PFES
(Panchuk et al. 1997), NES (Panchuk et al. 2006) spectrometers
($\lambda \lambda$ 4470--7100 \AA).
\end{enumerate}

The data reduction was done using IRAF and MIDAS software packages,
all the $RV$ measurements were done using the DECH20
software (Galazutdinov 1992). In Table 1 we present these $RV$
data from the spectra obtained in 2005--2010. This table
contains the measurements derived from the metallic
lines, H$_{\alpha}$, and H$_{\beta}$, respectively.\\[2mm]

\begin{table}[t]
\begin{center}
\caption{Radial velocity data of Polaris Aa in 1994--2010}
\small
\tiny
\begin{tabular}{rclccccc}
\hline
Spect-    &    HJD       & Tel.        & \multicolumn{3}{c}{Metallic lines} & H$_{\alpha}$ & H$_{\beta}$\\
rum  ID   & 2400000+ & & $RV$ & $\sigma$ & NL & $RV$ &  $RV$ \\
\hline
940609 & 49512.615 & 1      & -13.28 & 1.23 & 126 & -14.78 &    -   \\
940815 & 49579.824 & 1      & -14.21 & 1.21 & 116 & -15.53 &    -   \\
940908 & 49603.853 & 1      & -13.35 & 0.93 & 152 &    -   & -14.34 \\
941012 & 49637.792 & 1      & -14.97 & 1.05 & 132 & -16.69 &    -   \\
941023 & 49648.810 & 1      & -14.38 & 1.07 & 130 & -15.53 &    -   \\
s22923 & 51240.612 & 3      & -18.26 & 2.81 & 302 & -19.98 &    -   \\
s23908 & 51360.538 & 3      & -16.51 & 2.36 & 317 & -16.48 &    -   \\
s24008 & 51361.536 & 3      & -14.53 & 2.68 & 275 & -15.33 & -16.82 \\
011009 & 52192.858 & 2      & -16.88 & 0.81 & 281 & -19.23 &    -   \\
020522 & 52416.655 & 1      & -16.53 & 1.17 & 138 & -18.18 &    -   \\
020523 & 52417.616 & 1      & -17.67 & 1.45 & 145 & -19.35 &    -   \\
020527 & 52421.679 & 1      & -17.85 & 1.32 & 109 & -19.46 &    -   \\
020601 & 52426.667 & 1      & -18.18 & 1.28 & 121 & -20.39 &    -   \\
020602 & 52427.650 & 1      & -17.35 & 3.06 & 119 & -18.33 &    -   \\
020610 & 52435.634 & 1      & -16.53 & 1.18 & 142 & -19.19 &    -   \\
020616 & 52441.673 & 1      & -16.78 & 1.08 & 112 & -20.35 &    -   \\
s36713 & 52514.575 & 3      & -20.39 & 0.92 & 270 & -22.54 &    -   \\
s36814 & 52515.588 & 3      & -15.33 & 0.86 & 396 &    -   & -15.77 \\
s40008 & 52782.543 & 3      & -16.62 & 0.60 & 374 &    -   & -16.41 \\
031013 & 52833.741 & 1      & -21.64 & 1.30 &  93 & -14.25 &    -   \\
031017 & 52837.678 & 1      & -21.59 & 6.07 & 104 & -17.13 &    -   \\
031019 & 52839.746 & 1      & -23.97 & 4.00 & 111 & -15.07 &    -   \\
s40410 & 52861.560 & 3      & -17.76 & 0.73 & 279 & -20.47 &    -   \\
s40819 & 52867.562 & 3      & -17.75 & 0.79 & 251 & -20.25 &    -   \\
s40921 & 52869.570 & 3      & -16.62 & 0.76 & 247 & -19.08 &    -   \\
s41209 & 52891.600 & 3      & -16.38 & 0.89 & 384 &    -   & -15.76 \\
031109 & 52952.700 & 1      & -19.19 & 1.37 &  90 &  -7.46 &    -   \\
0312131& 52986.692 & 1      & -18.48 & 1.89 & 125 & -10.53 &    -   \\
0312132& 52986.709 & 1      & -17.86 & 1.67 & 107 &  -9.31 &    -   \\
040101 & 53005.595 & 1      & -16.50 & 1.13 & 141 &  -8.17 &    -   \\
s42006 & 53015.167 & 3      & -17.79 & 0.88 & 279 & -19.81 &    -   \\
s42202 & 53019.108 & 3      & -17.28 & 0.82 & 266 & -19.22 &    -   \\
s42302 & 53072.165 & 3      & -17.81 & 0.64 & 251 & -20.09 &    -   \\
s42327 & 53072.631 & 3      & -18.02 & 0.77 & 291 & -20.17 &    -   \\
s42421 & 53073.622 & 3      & -17.52 & 0.80 & 278 & -19.69 &    -   \\
s42502 & 53131.194 & 3      & -18.21 & 0.98 & 549 & -18.40 &    -   \\
s43302 & 53246.192 & 3      & -16.50 & 0.73 & 281 & -19.22 &    -   \\
s43812 & 53285.167 & 3      & -17.08 & 0.85 & 304 & -19.06 &    -   \\
041227 & 53367.091 & 2      & -20.51 & 3.84 & 261 & -23.39 &    -   \\
s45233 & 53686.647 & 3      & -17.68 & 1.05 & 198 &    -   & -17.54 \\
s45328 & 53687.637 & 3      & -15.82:& 1.00 & 616 &    -   & -16.48 \\
s45531 & 53689.649 & 3      & -18.24 & 1.20 & 589 & -18.60 &    -   \\
s45602 & 53690.111 & 3      & -17.80 & 1.13 & 566 & -17.50 &    -   \\
s45821 & 53691.635 & 3      & -17.82 & 1.06 & 550 & -17.34 &    -   \\
s45902 & 53693.124 & 3      & -17.93 & 1.06 & 549 & -18.41 &    -   \\
s463002& 53751.123 & 3      & -16.83 & 1.21 & 581 &    -   & -16.72 \\
s466002& 53808.277 & 3      & -18.78 & 1.55 & 933 &    -   & -19.43 \\
s469012& 53904.348 & 3      & -17.87 & 1.09 & 506 &    -   & -17.21 \\
s478030& 53980.588 & 3      & -17.40 & 1.29 & 569 & -17.20 &    -   \\
s482001& 54073.591 & 3      & -18.43 & 1.15 & 579 & -17.88 &    -   \\
s485029& 54077.653 & 3      & -17.58 & 1.21 & 406 &    -   & -17.27 \\
s494030& 54169.639 & 3      & -19.18 & 1.09 & 415 &    -   & -20.09 \\
s497012& 54225.226 & 3      & -18.92 & 1.25 & 592 & -18.59 &    -   \\
s504049& 54344.551 & 3      & -19.41 & 1.04 & 464 & -18.47 &    -   \\
s510001& 54426.185 & 3      & -16.65 & 1.19 & 603 & -16.26 &    -   \\
s532015& 54934.587 & 3      & -17.19 & 1.10 & 573 & -16.61 &    -   \\
\hline
\end{tabular}
\begin{list}{}{}
\item Column 1 lists the spectrum ID shown in Fig.\,1; column 3 shows telescopes used (see below);
columns 4 and 5 show the average RV and r.m.s. error from the metallic lines; column 6
lists the number of metallic lines (NL) measured.
\item[1] -  1\,m Ritter Observatory;
\item[2] - 2.1\,m McDonald Observatory;
\item[3] - 6\,m Special Astrophysical Observatory, Russian Academy of
Sciences.
\end{list}
\end{center}
\end{table}

\begin{figure}[h]
\resizebox{\hsize}{!}{\includegraphics{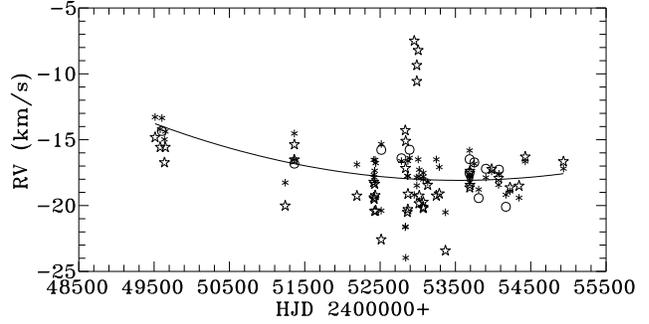}}
\caption{Radial velocity measurements in the spectra of Polaris Aa during 1994--2010. Six-point
stars -- from the metallic lines, open five-point stars -- from the H$_{\alpha}$
line, and open circles -- from the H$_{\beta}$ line.  The solid line shows a second degree polynomial approximation
for the metallic lines.}
\label{Fig1}
\end{figure}

{\bf 2. Radial velocity measurement analysis and the H$_{\alpha}$ line core
behaviour}\\[1mm]

As seen in Table 1 and Fig.\,1, the difference between the
measurements obtained from the metallic lines and H$_{\alpha}$ (and one from H$_{\beta}$)
for each spectrum does not exceed 1.5 \kms in 1994. As seen from
Fig. 2, the H$_{\alpha}$ core does not demonstrate any visible asymmetries.

Since 1999 (HJD 2451240--2451361) this difference begins to increase
(Fig. 1) and a slight asymmetry on the red side of the H$_{\alpha}$
core are visible (Fig. 2). Two years later this difference becomes larger
(from 1 \kms to 2 \kms), and the asymmetries on the red side of the core get quite
visible (Fig. 3) during two years (2001--2002).

During 2003 one can see the most interesting event when the difference
between the measurements reaches 8--12 \kms (see Table 1 and Fig. 1) and
the H$_{\alpha}$ core shows asymmetries on the red side as well as on the blue side
(see Fig. 4).

Since 2004 this difference decreases to 2--2.5 \kms (HJD 2453015--2453367),
and the H$_{\alpha}$ core exhibits asymmetries on the red side only (see Fig. 1 and 5).

During 2005--2006 (HJD 245689--2454073) the difference is less than 1 \kms and
the asymmetries are less visible (Fig. 6). The same phenomena one can see in other
results obtained during 2008--2010 (HJD 2454077--2454934) (Fig. 7). It
should be noted that the differences between the H$_{\alpha}$ and H$_{\beta}$ line measurements
are negligible.\\

{\bf 3. Conclusions}\\[1mm]

We summarize the results of our investigation as follows.

\begin{enumerate}
\item As seen from the results in Table 1 and Fig.1, amplitudes of the $RV$ curve
from H$_{\alpha}$ and H$_{\beta}$ are very small and close to those
determined from the metallic lines.
\item First H$_{\alpha}$ line core asymmetries on the red side arise
with an increase of the $RV$ curve amplitude after the historical minimum of
the Polaris Aa pulsational activity in the beginning of the 1990's.
\item During 2003 the difference between the metallic line $RV$ and those from the
H$_{\alpha}$ line core reaches 8--12 \kms. This event is accompanied by
the pronounced asymmetries of the H$_{\alpha}$ core on the red side as well as on the
blue side.
\item Since 2004 the H$_{\alpha}$ line core asymmetries are observed on the red side
only and nearly disappear after 2005, when the $RV$ amplitude grows to a new
minimum.
\item H$_{\alpha}$ core asymmetries (so-called a knife-like profile) in the atmosphere of
Polaris Aa show that this absorption line could be an indicator
of the envelope presence in yellow pulsating supergiants with short periods
and small amplitudes.
\item So unusual behaviour of the H$_{\alpha}$ core during 2003 could be explained
by dynamical changes in the envelope around Polaris Aa.[2mm]
\end{enumerate}

{\bf 3. Acknowledgments}\\[1mm]

This study was financially supported by the SCOPES Swiss National Science Foundation (project no. IZ73Z0152485).\\[2mm]

{\bf References\\[2mm]} \noindent Galazutdinov, G.A.: 1992, {\it
Preprint SAO RAS} No {\bf 92}, 1\\
McCarthy, J.K., Sandiford,
B.A., Boyd, D. \& Booth, J.: 1993, {\it PASP} {\bf 105}, 881\\
M\'{e}rand, A., Kervella, P., Coud\'{e} du Foresto, V., Perrin, G.
et al.: 2006, {\it A\&A} {\bf 453}, 155\\
Panchuk, V.E.,
Klochkova, V.G., Galazutdinov, G.A., Radchenko, V.P., Chentsov,
E.L.: 1993, {\it AstL}~ {\bf 19}, 431\\
Panchuk, V.E., Najdenov,
I.D., Klochkova, V.G., Ivanchik, A.B., Yermakov, S.V., Murzin,
V.A.: 1997, {\it Bull. SAO RAS} {\bf 44}, 127\\
Panchuk, V.E.,
Klochkova, V.G., Nadenov, I.D., Yushkin, M.V.: 2006, {\it The
Ultraviolet Universe: Stars from Birth to Death, 26th Meeting of
the IAU, Joint Discussion 4, 16-17 Aug. 2006, Prague, Czech
Republic}~ {\bf D04}, 14\\
Usenko, I.A., Kniazev, A.Yu.,
Berdnikov, L.N., Kravtsov, V.V., Fokin, A.B.: 2013, {\it AstL}
{\bf 39}, 432\\
Usenko, I.A., Kniazev, A.Yu., Berdnikov, L.N.,
Fokin, A.B., Kravtsov, V.V.: 2014a, {\it AstL} {\bf 40}, 435\\
Usenko, I.A., Kniazev, A.Yu., Berdnikov, L.N., Kravtsov, V.V.:
2013b, {\it AstL} {\bf 40}, 800\\
Usenko, I.A., Klochkova, V.G.:
2015 {\it AstL} {\bf 41}, 351

\begin{figure}[t]
\resizebox{\hsize}{!}{\includegraphics{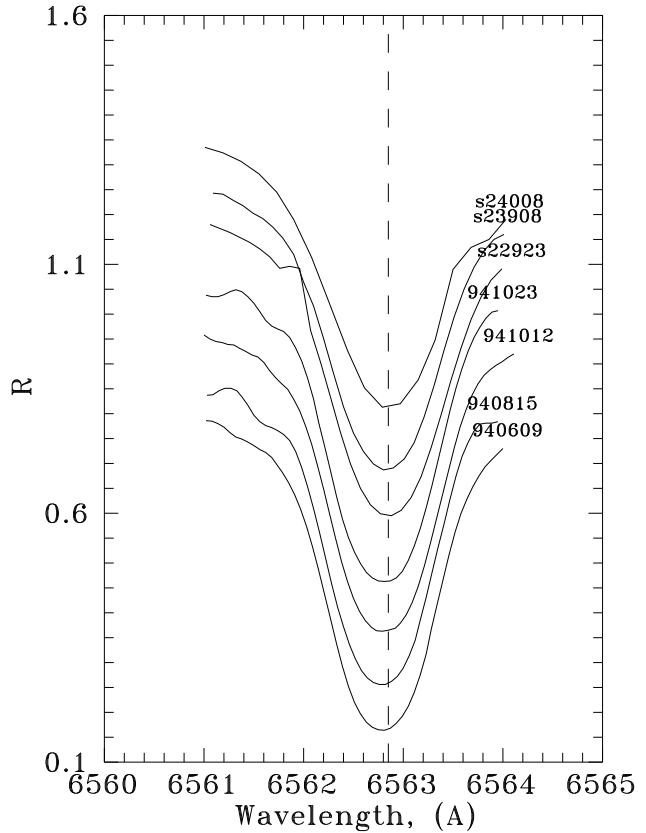}}
\caption{The H$_{\alpha}$ line core profiles of Polaris Aa in 1994--1999. The intensity ($R$) is shown in the
units of the underlying continuum, the wavelengths are shown in Angstr\"oms.}
\label{Fig2}
\end{figure}


\begin{figure}
\resizebox{\hsize}{!}{\includegraphics{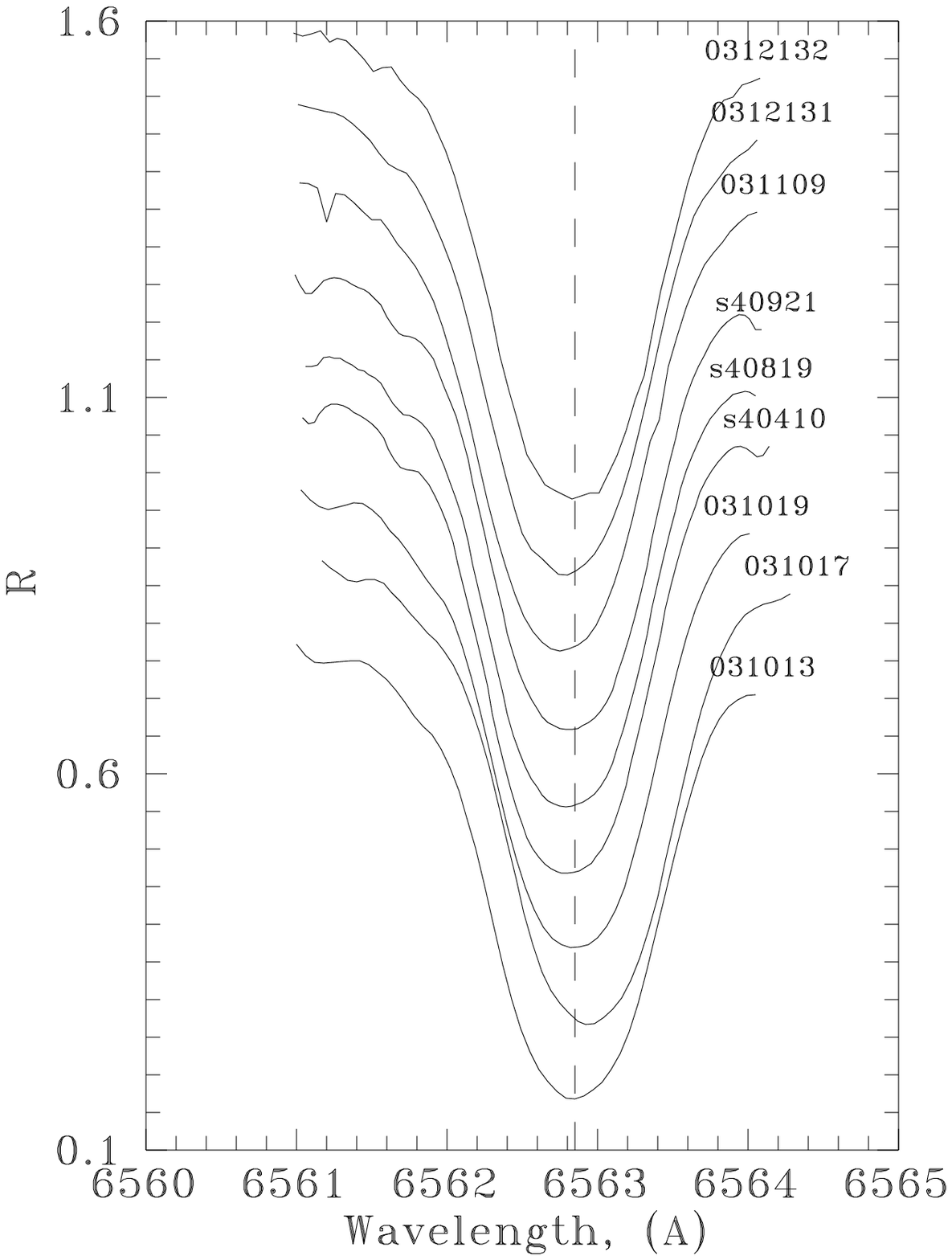}}
\caption{The H$_{\alpha}$ line core profiles of Polaris Aa in 2003.}
\label{Fig3}
\end{figure}

\begin{figure}
\resizebox{\hsize}{!}{\includegraphics{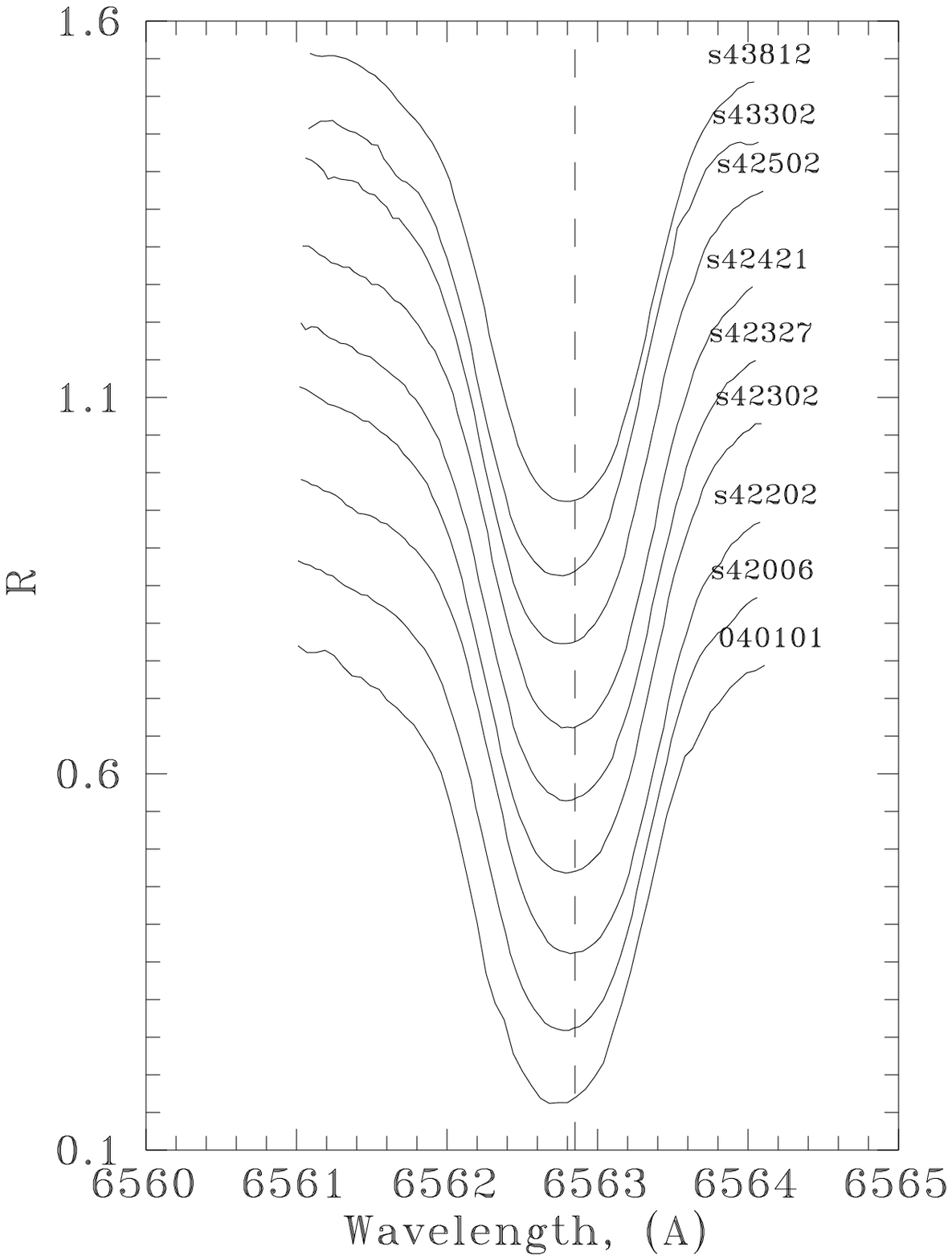}}
\caption{The H$_{\alpha}$ line core profiles of Polaris Aa in 2004.}
\label{Fig4}
\end{figure}

\begin{figure}
\resizebox{\hsize}{!}{\includegraphics{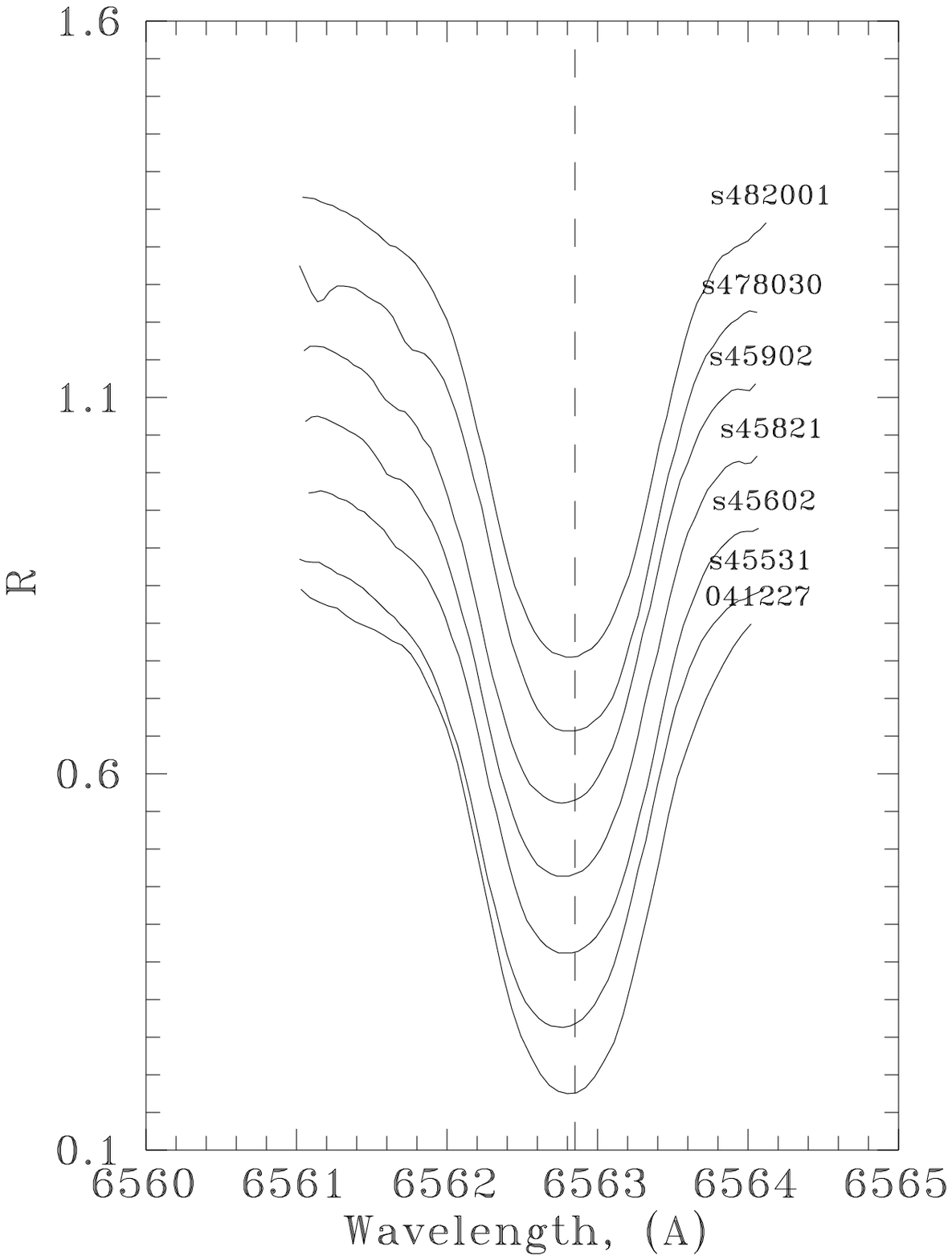}}
\caption{The H$_{\alpha}$ line core profiles of Polaris Aa in 2005--2006.}
\label{Fig5}
\end{figure}

\begin{figure}
\resizebox{\hsize}{!}{\includegraphics{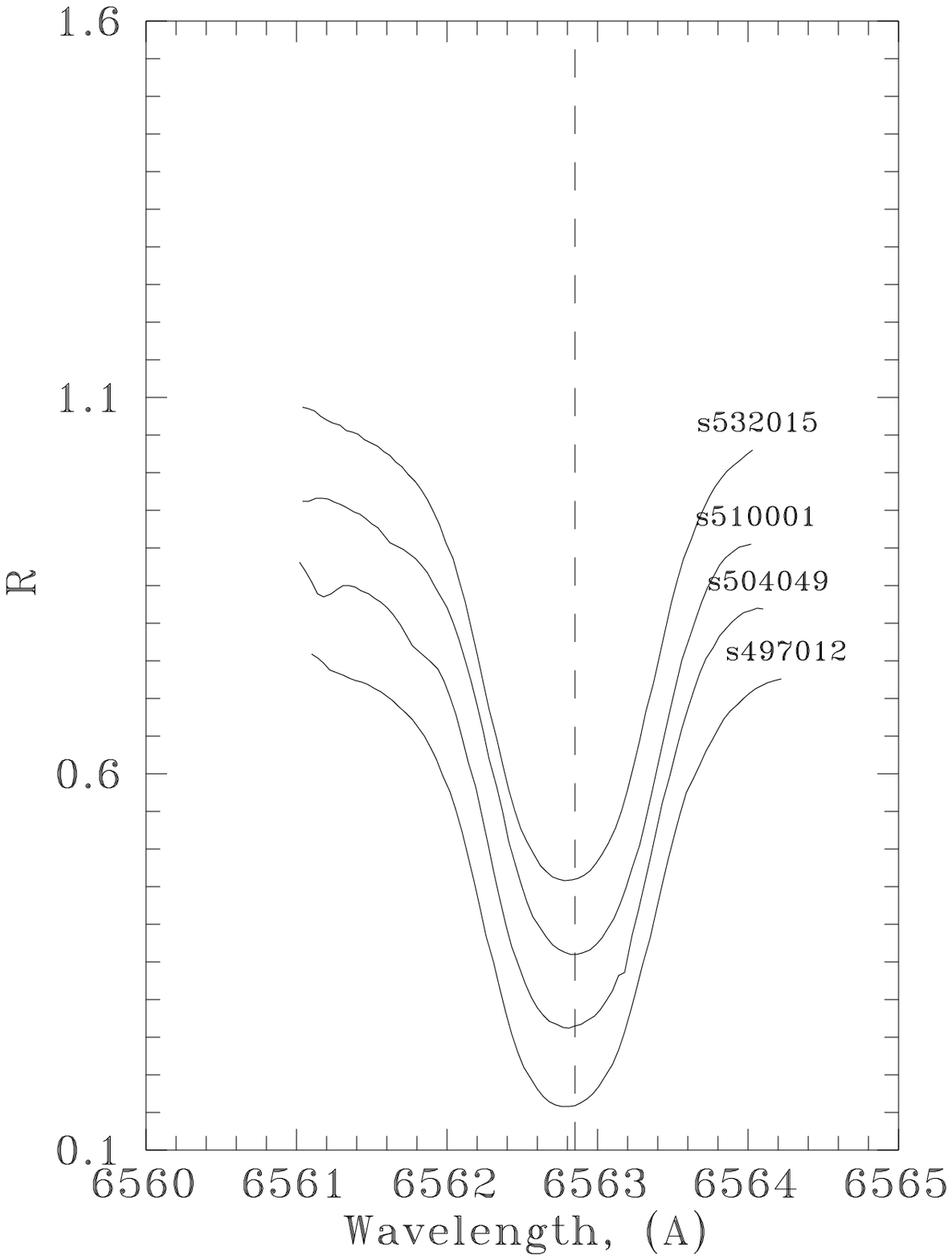}}
\caption{The H$_{\alpha}$ line core profiles of Polaris Aa in 2008--2010.}
\label{Fig6}
\end{figure}

\end{document}